\documentclass[aps,showpacs,prl,twocolumn,groupedaddress]{revtex4}

\usepackage{graphicx,amssymb,amsmath}
\usepackage{natbib}

\newcommand{\gd}{\dot{\gamma}}

\newcommand{\D}{\Delta}

\begin{document}

\bibliographystyle{apsrev}


\title[]{Shear-induced Diffusion in Dilute Suspensions of Charged Colloids}

\author{Victor Breedveld}
\email[]{victor@engineering.ucsb.edu}

\author{Alex J. Levine}
\email[]{ajlevin@mrl.ucsb.edu}

\affiliation{Department of Chemical Engineering \& Materials
Research Laboratory, University of California at Santa Barbara,
Santa Barbara, CA 93106.}

\date{\today}

\begin{abstract}
We propose a model for the nonequilibrium enhancement of colloidal
self--diffusion in an externally imposed shear flow in charged
systems. The enhancement of diffusion is calculated in terms of
the electrostatic, two--body interactions between the particles in
the shear flow.  In the high shear rate, low volume fraction limit
in which our model is valid, we compare these calculations to the
experiments of Qiu {\it et al.\/} [Phys. Rev. Lett. \textbf{61}, 2554 (1988)].

\end{abstract}

\maketitle

Colloidal suspensions \cite{Murray&95,Russel&89} serve as a unique model
system for both atomic liquids and solids. Due to their
inherently longer length and time scales, colloidal systems offer
the opportunity to probe the many--body, nonequilibrium dynamics
of interacting systems through the application of {\it e.g.\/} moderate
shear. Such issues involving systems driven far from equilibrium remain
among the principal puzzles of modern statistical mechanics.

In this letter we discuss the origin of a particularly intriguing
and apparently generic feature of particulate suspensions driven
out of equilibrium by the application of shear flow.  Since its
initial observation by Qiu {\it et al.\/} \cite{Qiu&88} in a
charged colloidal suspension, it is now well--known that the
effective single--particle diffusion constant grows under shear.
Similar enhancement of self--diffusion under applied shear been
observed in concentrated non-colloidal ({\it i.e.\/} non-Brownian or
high Peclet number) suspensions both experimentally \cite{Leighton&87a,Phan&98,Breedveld&98}
and in simulations \cite{Foss&99}.  It should be pointed out that the diffusion
enhancement under consideration occurs in the gradient and vorticity
directions, and is thus unrelated to the better understood Taylor
dispersion which contributes to the effective diffusion
along the flow direction.

Since laminar, zero Reynolds number flows obey time reversal symmetry,
one is forced to look elsewhere for the symmetry breaking required
for understanding the phenomenon of nonequilibrium diffusion enhancement.
In uncharged,
non-colloidal suspensions it has been shown numerically that
three-particle hydrodynamic interactions are chaotic
\cite{Wang&96,Janosi&97}. These chaotic interactions have been
previously implicated as a source of non-thermal noise in
nonequilibrium suspensions \cite{Levine&98}. In addition, finite
particle roughness \cite{Cunha&96} can break the
symmetry of low Reynolds number hydrodynamics effectively by
cutting off lubrication forces at small interparticle separations.

\begin{figure}[htpb]
  \centering
  \includegraphics[width=8.0cm]{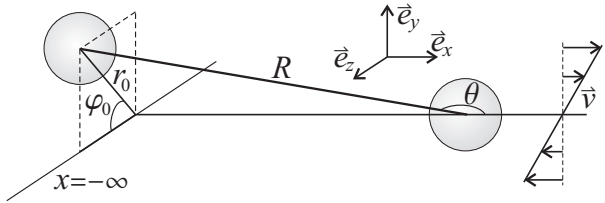}
  \caption{Coordinate system for two-particle collision in a shear flow;
  the right particle is located at the origin of the frame of reference.}
  \label{fig:coord}
\end{figure}

In interacting colloidal systems there is, however, another source
of symmetry breaking due to the electrostatic interactions between
the colloids \cite{Russel&89,DLVO} (See Fig.~\ref{fig:coord}).  The importance of the
electrostatic interactions in the diffusion enhancement observed
by Qiu {\it et al.\/} is demonstrated by the fact that a decrease
in the Debye screening length suppresses the effect. We focus
exclusively on this experiment and  the role of the electrostatic
interactions since these interactions are more analytically
tractable than the chaotic, many--body hydrodynamics that
presumably generates a similar diffusion enhancement in
non-Brownian hard--sphere suspensions.

Our calculation proceeds as follows:  We consider the
randomization of the trajectory of a particle, which we refer to
as the scattering center, due to its electrostatic interaction
with a spatially random collection of identical particles that are
carried past it in the shear flow. This is done in two steps.  We
first compute the trajectory of a charged colloid as the
macroscopic shear flow carries it past the scattering center. From
this calculation we determine the displacement of the scattering
center itself.

Given a spatially random ensemble of such particles being carried
by the scattering center by the simple shear flow (uniform rate of shear
equal to $\gd$), we compute the second moment of the
displacements of the scattering center due to interactions
with the particles flowing past it. The product of this second
moment with the rate of scattering events ($\sim \gd$) gives the
shear enhancement of the diffusion constant. The result of this
calculation is the diffusion enhancement: $\Delta D \sim \left|
\gd \right|^\alpha$.  We compare both the magnitude of this
diffusion enhancement and its dependence upon shear rate ({\it
i.e.\/} $\alpha$) with experiment.

\begin{figure}[htpb]
  \centering
  \includegraphics[width=6.0cm]{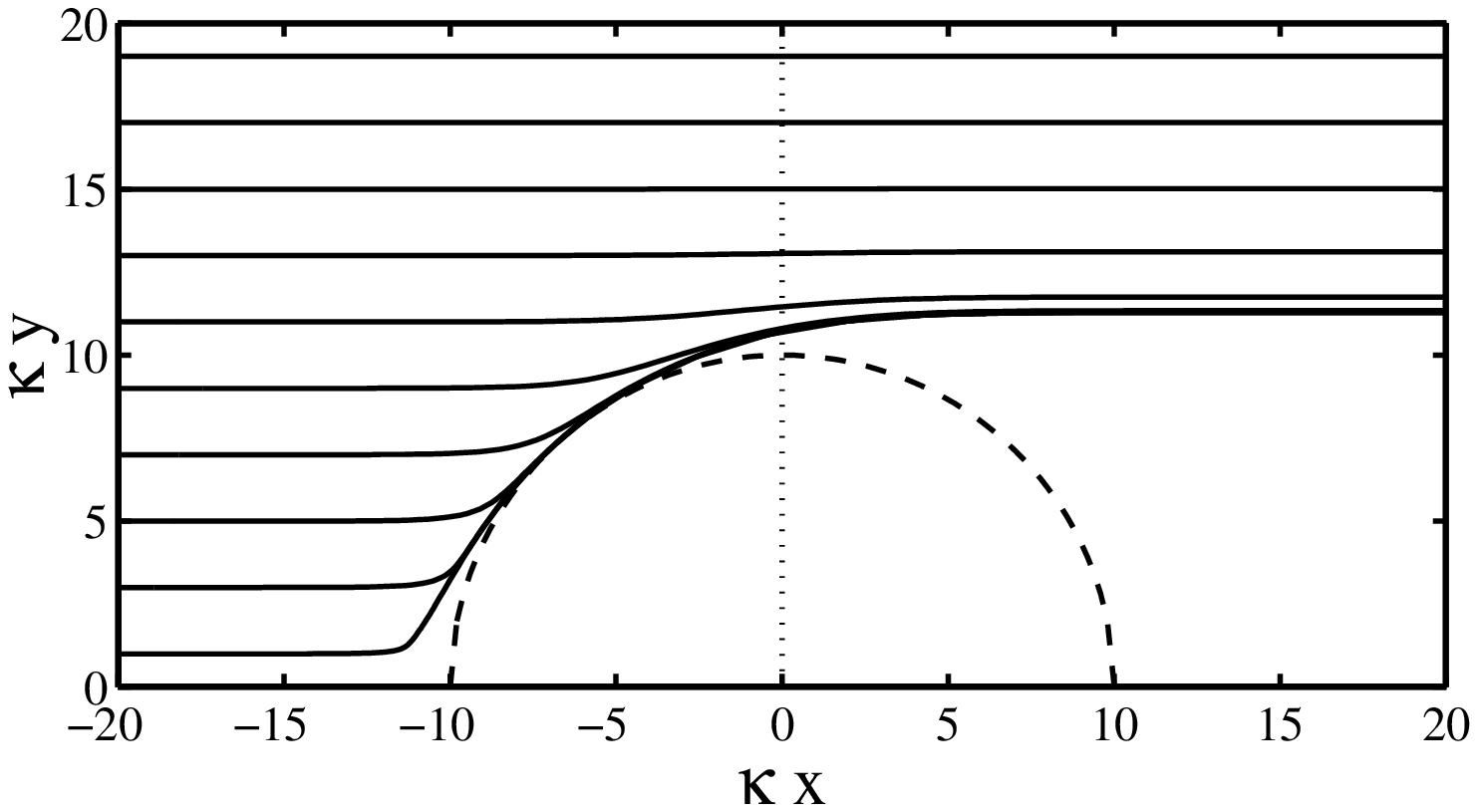}
  \includegraphics[width=6.0cm]{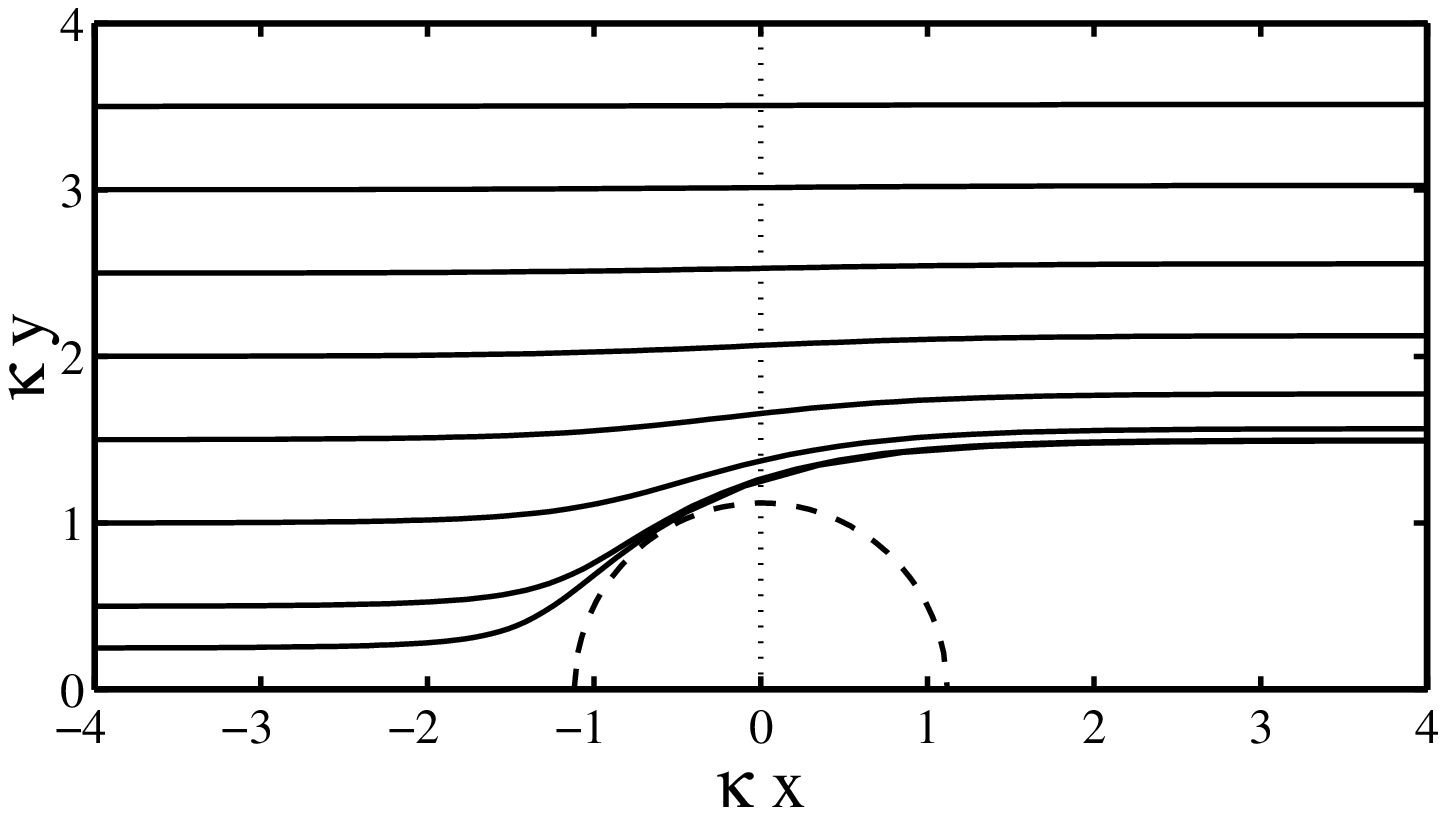}
  \caption{Trajectories in the $x-y-$plane (\textit{i.e.} $\varphi_0=\pi/2$)
  for the collision between two electrostatically interacting colloids
  in a shear flow; $H=10^6$ in the upper figure and $H=1$ in the lower figure;
  trajectories are plotted for different impact parameters $\hat{r}_0$.
  The dashed semicircles denote the volume of the interaction
  zone surrounding the scattering center.}
  \label{fig:traj}
\end{figure}

Qualitatively, the trajectory of a charged particle under the
combined influence of the macroscopic shear flow and the
electrostatic interaction with the scattering center can be
described as follows: The shear flow brings a particle toward the
scattering center along a given stream line. While the particle is
within a few screening lengths of the scattering center, the
electrostatic interaction displaces the particle from its stream
line.  However, once the particle has been carried by the
combination of flow and electrostatic interaction to a distance of
more than a few screening lengths, the particle resumes its
trajectory along a {\it different} streamline than the one it
entered on. See Fig.~\ref{fig:traj} for computed trajectories.

There are two important physical limits which result in
qualitatively different particle trajectories and a different
dependence of diffusion enhancement upon shear rate.  The limits
correspond to the dominance of one of two independent forces
controlling the particle trajectories, hydrodynamic drag and the
interparticle, electrostatic interaction.

In the case where the electrostatic interaction is arbitrarily
strong so that it completely controls the motion of a particle in
the interaction zone, we recover a simple linear dependence of the
diffusion enhancement with shear. A particle enters the
interaction zone along some stream line. Now the dominant
electrostatic repulsion between that particle and the scattering
center drives the particle out of the interaction zone. It then
resumes its straight--line motion along a {\it different\/}
streamline. The displacement of the particle is thus on the order
of the interaction zone radius and independent of the imposed
shear rate. The scattering rate, however, is proportional to the
shear rate, so one the diffusion enhancement scales linearly with
shear rate, $\Delta D \sim \left| \gd \right|$ in the low shear
rate regime where the particle--particle interaction is entirely
dominated by the electrostatics. See Fig. 2a.

On the other hand, if the electrostatic interaction is weak
enough, or, equivalently, if the shear rate is high enough, a
particle will be carried predominantly by the shear flow through
the electrostatic interaction zone surrounding the scattering
center.  In this case the residence time of the particle in the
interaction zone is inversely proportional to the shear rate.  In
the low Reynolds number limit (which we always assume) the
displacement of the particle from its initial stream line is
proportional to the time integral of the force acting on it.  In
this case that time integral will be proportional to $\gd^{-1}$.
Since the rate of scattering events will still be proportional to
$\gd$, we expect that the enhancement of diffusion will plateau at
high shear rates.  See Fig. 2b.

The above, heuristic cases describe only the limiting cases of
strong and weak shear. Not surprisingly, for physical values of
the ratio of these two forces including those encountered in
experiment, the predicted result for the shear enhancement
interpolates between these two limits.  We find for a broad range
of shear rates and for physically relevant electrostatic
interaction parameters that the shear enhancement of diffusion
scales as $\Delta D \sim \left| \gd \right|^{0.7}$.

Our calculation cannot be extended to zero $\gd$. We estimate a
lower shear--rate cutoff for our analysis by noting that we assume
that particles follow stream lines towards and away from the
scattering center except when they are within the interaction zone
surrounding each particle.  If the shear rate is so low that
particles can diffuse through the interaction zone, the
deterministic trajectory calculations are no longer valid.The
cross--over from our model calculations to quadratic scaling
occurs at a critical shear rate $\dot{\gamma}^\star \sim \kappa^2
D_0$ where $D_0$ is the Brownian diffusion constant. At this
cross-over shear rate a particle diffuses a Debye length in the
same time as it is advected that distance by the flow.  For the
experimental system in question $\dot{\gamma}^\star =
10^{2}\;\mathrm{s^{-1}}$.  In addition, we note that the reduction
of many--body interactions into a series of simple two--body
interactions can only be justified in the limit of small particle
number density.

This result for the scaling of the shear enhancement of diffusion
with shear rate is in good agreement with nonequilibrium,
Brownian--dynamics simulations \cite{Xue&89,Chakrabarti&94}. The
experimental data \cite{Qiu&88}, on the other hand, is consistent
with a power law of unity.  To resolve this discrepancy, we point
out that our calculation is not applicable to the lower shear rate
experimental data.  In this lower shear rate regime, we expect
that the shear--rate dependence of $\Delta D$ is {\it stronger\/}
than linear\cite{Morris&96} so that the effect of the cross--over
of the exponent from greater than one to  $0.7$ leads to a larger
apparent exponent in the data\cite{footnote-Indrani}. Secondly, we
note that, over the single decade of shear--rate data available,
it is problematic to distinguish between our proposed exponent of
$0.7$ and the linear dependence on shear rate concluded by Qiu
{\it et al.\/}. In addition, we will show, that by making
reasonable assumptions about the electrostatic interactions
present in the experimental system of Qiu {\it et al.\/}, our
calculation of the magnitude of the shear enhancement effect
agrees well with the experiment.

In an earlier mode--coupling study Indrani {it et al.\/}
\cite{Indrani&95} attribute the diffusion enhancement to the
suppression of the effective friction, $\zeta$, experienced by a
concentration fluctuation. Noting that $D \sim T/\zeta$ they
calculate the contribution to $\zeta$ from the interaction of a
particle concentration fluctuation with all other thermally
generated concentration fluctuations in the medium.  The principal
effect of the shear flow is to destroy these concentration
fluctuations and thereby decrease the effective friction. We, on
the other hand, propose that the principal effect of the shear
flow when considered along with the electrostatic interparticle
interactions is to increase the effective temperature in the
nonequilibrium system.

The particle trajectories are described by the following
differential equation \cite{Vandervorst&97}:
\begin{equation}\label{eq:diffeq}
\frac{dY}{d\xi}=Y\frac{\xi}{1+\xi^2}+\frac{H}{\sin\varphi}\frac{Y+1}{Y^2}e^{-Y}
\end{equation}
where $Y=\kappa R$ is the distance of the incoming particle from
the scattering center measured in Debye lengths $\kappa^{-1}$ and
$\xi=\cot\theta$ is the polar angle. See Fig.~\ref{fig:coord}.
Eq.~\ref{eq:diffeq} accounts for the interplay of the hydrodynamic
drag force by the fluid represented by the first term on the LHS
of and electrostatic repulsion between particles represented by
second term on the LHS of Eq.~\ref{eq:diffeq}. The relative
importance of these two forces is measured by the dimensionless
parameter $H$, which represents the ratio between the
electrostatic interparticle interaction and the hydrodynamic drag.
$H$ is defined by
\begin{equation}\label{eq:H}
 H=\frac{8 \pi \epsilon_r \epsilon_0 \Phi_a^{2} (\kappa a)^2 e^{2 \kappa a
 }\kappa}{\zeta\:\gd}=\frac{C}{\gd},
\end{equation}
where $\epsilon_r\epsilon_0$ is the dielectric constant of the
fluid, $\Phi_a$ the apparent surface potential of the particles,
and $\zeta$ the hydrodynamic friction factor. All the details of
the electrostatic interaction for a given colloidal system can be
subsumed into the parameter $C$ defined above.

Eq.~\ref{eq:diffeq} neglects hydrodynamic interactions between
particles which is acceptable if particles do not make close
passes. In a dilute suspension of highly charged colloids the
number of such close passes should be negligible and thus not
greatly contribute to the effective diffusion constant.  This view
is supported by the original experiments which showed that the
shear-enhancement of diffusion disappears with sufficient
screening of the electrostatic interaction \cite{Qiu&88}.

To determine the trajectory of a scattering particle we integrate
Eq.~\ref{eq:diffeq} from initial conditions such that at large
distances from the scattering center ({\it i.e.\/} $x
\longrightarrow - \infty$) the incoming particle has an impact
parameter $r_0$ and has a polar angle $\varphi_0$ measured from
the $\hat{z}$--axis in the $yz$ plane. As a result of the
azimuthal symmetry about the $\hat{x}$--axis, every trajectory is
confined to a plane of constant $\varphi$.  However, because
$\hat{y}$ is the velocity gradient direction of the imposed shear
flow, the rate of incoming particles to the scattering center will
depend on $\varphi$.  We show in Fig.~\ref{fig:traj} two
numerically integrated trajectories.  The first shows the result
obtained in the electrostatically dominated regime (high $H$) and
the second shows the shear flow dominated regime (low $H$).

We determine the total displacement of the in--coming particle
normal to the streamlines by finding the change in the radial
distance $\Delta \hat{r} = \kappa (r|_{x \longrightarrow \infty} -
r_0$) of the particle's trajectory from the $\hat{x}$--axis
comparing the asymptotic out--state where $x \longrightarrow
\infty$ to the initial impact parameter.  By symmetry, each
particle (the incoming particle and the scattering center) moves
half of the distance $\Delta r$ in the laboratory frame.

The effective enhancement of the diffusion constant in both the
velocity gradient ($y$) and vorticity ($z$) directions can be
calculated from the second moment of the particle displacements in
the appropriate directions:
\begin{eqnarray}
\label{eq:Dyy1}
 \Delta D_{yy}=\tfrac{1}{2} \kappa^{-2} \int_{s} P(s)\;\left(\tfrac{1}{2}\:\D \hat{r}(s)\:\sin\varphi_0\right)^2 d^{2}s  \\
 \label{eq:Dzz1}
\Delta D_{zz}=\tfrac{1}{2} \kappa^{-2} \int_{s}
P(s)\;\left(\tfrac{1}{2}\:\D \hat{r}(s)\:\cos\varphi_0\right)^2 d^{2}s
\end{eqnarray}
where $P(s)\:ds$ is the probability per unit time of a collision
to occur with impact parameter $s=(\hat{r}_0,\varphi_0)$, where $\hat{r}_0=\kappa r_0$. The
terms in the parentheses are the displacements of a given particle
in the respective directions. Assuming a random distribution of
particles, $P(s)$ is given by the particle flux through the area
$r_0\:dr_0\:d\varphi_0$:
\begin{equation}\label{eq:P}
P(s)\;d^{2}s=\frac{3\:\phi}{4\pi\:(\kappa a)^3}\gd\;\sin\varphi_0\;\hat{r}_0^2\:d\hat{r}_0\:d\varphi_0
\end{equation}
The integration domain in Eqs.~\ref{eq:Dyy1},\ref{eq:Dzz1} only
accounts for particles flowing in from the left ($y>0$).  The
values for the elements of the diffusion tensor have been doubled
to account for the contribution of the particle flux coming from
the right ($y<0$). As expected, the shear rate $\gd$ affects
shear-enhanced diffusion in two distinct ways: through the
particle flux ($\sim \gd$) and the step size, which depends on
$H=C/\gd$ in a complicated way that we explore below.

To explore the shear rate dependence of the diffusion enhancement,
it is convenient to express the diffusion constants in the
following form using Eqs.~\ref{eq:Dyy1},\ref{eq:Dzz1}, and
\ref{eq:H}:
\begin{equation}
\label{eq:Idef}
 \Delta D_{\alpha \alpha} = \frac{3 \phi }{4 \pi (a
\kappa)^3 \kappa^2} \frac{C}{H} I_\alpha(H),
\end{equation}
where $I_\alpha$ is a dimensionless integral and $\alpha = y,z$.
The entire shear rate dependence of the two diffusivities is
contained in $1/H \cdot I_{\alpha}(H)$ where $H \sim 1/\gd$.
For small values of $H$ (large $\gd$ or low interaction), the $H$
dependence of $I_\alpha(H)$ is linear as can be seen by simple
perturbation theory in $H$. $\Delta D_{\alpha \alpha}$ at these
large shear rates becomes independent of $H$. For large $H$,
however, the dependence on $H$ weakens and the curves are
described well by $I_{\alpha}(H)~\sim H^{0.3}$ so that the diffusion
enhancement, which scales as $1/H\cdot I_\alpha(H)$ takes the form:
$\Delta D_{\alpha} \sim 1/H \cdot I_\alpha(H) \sim H^{-0.7} \sim
\gd^{0.7}$. That this power law holds over a wide range of shear
rates is one of the central results of this letter. Such power law
scaling is difficult to distinguish from an exponent of unity over
the range of shear rates where we expect the model to be valid
($100-1000\;\mathrm{s^{-1}}$)  The scaling is not inconsistent
with the experimental data of Qiu {\it et al.\/}. In addition, the
computed exponent agrees well with previous simulations of the
system \cite{Chakrabarti&94}.

\begin{figure}[htpb]
  \centering
  \includegraphics[width=7.0cm]{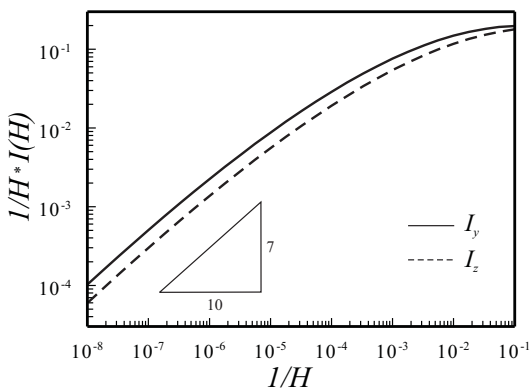}
  \caption{Scaling of diffusion with shear rate represented by the
  equivalent plot of $I(H)/H$ vs. $1/H$.}
  \label{fig:scale}
\end{figure}

The dependence of the  diffusivities upon $1/H$ demonstrating the
cross--over from the $0.7$ power law to the $H$--independence is
shown in Fig.~\ref{fig:scale}. This scaling of the diffusivity
with shear rate is distinct from that observed in non-colloidal
suspensions, where shear-induced diffusion grows monotonically
with $\gd$. At small shear rates the diffusion enhancement is anisotropic:
$\Delta D_{yy}/\Delta D_{zz}\approx 1.7$.

The calculated magnitude $\Delta D$ is also consistent with the
experimental data. To test this point, we need to find $C$. It is
determined from a combination of the electrostatic properties of
the suspension including the surface potential of the colloids
$\Phi_a$, the Debye screening length $\kappa^{-1}$, and the
particle radius as well as the inverse of the effective particle
mobility. Since we determine sphere mobility empirically from
single--particle diffusion constant of the uncharged system, we
have only to know the electrostatic parameters of the charged
system in order to determine $C$, and thus determine the
calculated diffusivity enhancement for the sheared suspension.

Unfortunately, Qiu {\it et al.\/} do not report a value for the
surface potential of the charged colloids. In spite of the fact
that the experimental system is not well enough characterized to
fix $C$, we can, by using the surface potential as a fitting
parameter, force $H$ to be in the range where we find the scaling
relation $\Delta D \sim \left|\gd \right|^{0.7}$. To achieve this
we set $\Phi_a = 200 \mbox{\rm mV}$, which is not atypical in such
charged colloidal systems. When $H$ is set to $10^5$ and $\gd$ is
chosen equal to be $500\;\mathrm{s^{-1}}$ ---in the middle of the
experimental range--- our model predicts $\Delta D_{zz} \sim 3
\cdot 10^{-12}\;\mathrm{m^2/s}$. This prediction is reasonably
close to the experimental result of $\Delta D_{zz} \sim 1 \cdot
10^{-12}\;\mathrm{m^2/s}$, considering the strong dependence of
$\Delta D$ on the electrostatic parameters, which are known
imprecisely at best (e.g. $\Delta D \sim
\kappa^{-4.1}\:e^{0.6\kappa a}$). The agreement between experiment
and theory demonstrates only the plausibility of the latter.
Diffusion measurements under shear need to be performed on better
characterized systems to rigorously test the theory. Furthermore,
systematic variation of the electrostatic parameters will provide
insight into the validity of our scaling predictions.

\begin{acknowledgments}
VB was supported by the Netherlands Organization for Scientific
Research (NWO).  AJL acknowledges support by the NSF under award
DMR-9870785. We thank D. Pine, S. Ramaswamy and A. Sood for useful and enjoyable discussion.
\end{acknowledgments}


\begin{thebibliography}{17}
\bibitem{footnote-Indrani}
  We note that similar arguments were made in ref
  \cite{Indrani&95}.
\end{thebibliography}
\end{document}